\documentclass[twoside,twocolumn,9pt]{article}
\usepackage{extsizes}
\usepackage{enumerate}
\usepackage[super,sort&compress,comma]{natbib} 
\usepackage[version=3]{mhchem}
\usepackage[left=1.5cm, right=1.5cm, top=1.785cm, bottom=2.0cm]{geometry}
\usepackage{balance}
\usepackage{mathptmx}
\usepackage{sectsty}
\usepackage{graphicx} 
\usepackage{lastpage}
\usepackage{subcaption}
\usepackage[format=plain,justification=justified,singlelinecheck=false,font={stretch=1.125,small,sf},labelfont=bf,labelsep=space]{caption}
\usepackage{float}
\usepackage{fancyhdr}
\usepackage{fnpos}
\usepackage[english]{babel}
\addto{\captionsenglish}{%
  
}
\usepackage{array}
\usepackage{droidsans}
\usepackage{charter}
\usepackage[T1]{fontenc}
\usepackage[usenames,dvipsnames]{xcolor}
\usepackage{setspace}
\usepackage[compact]{titlesec}
\usepackage{hyperref}
\usepackage{ulem}

\definecolor{cream}{RGB}{222,217,201}

\begin{document}

\pagestyle{fancy}
\thispagestyle{plain}
\fancypagestyle{plain}{
\renewcommand{\headrulewidth}{0pt}
}

\makeFNbottom
\makeatletter
\renewcommand\LARGE{\@setfontsize\LARGE{15pt}{17}}
\renewcommand\Large{\@setfontsize\Large{12pt}{14}}
\renewcommand\large{\@setfontsize\large{10pt}{12}}
\renewcommand\footnotesize{\@setfontsize\footnotesize{7pt}{10}}
\makeatother

\renewcommand{\thefootnote}{\fnsymbol{footnote}}
\renewcommand\footnoterule{\vspace*{1pt}%
\color{cream}\hrule width 3.5in height 0.4pt \color{black}\vspace*{5pt}} 
\setcounter{secnumdepth}{5}

\makeatletter 
\renewcommand\@biblabel[1]{#1}            
\renewcommand\@makefntext[1]%
{\noindent\makebox[0pt][r]{\@thefnmark\,}#1}
\makeatother 
\renewcommand{\figurename}{\small{Fig.}~}
\sectionfont{\sffamily\Large}
\subsectionfont{\normalsize}
\subsubsectionfont{\bf}
\setstretch{1.125} 
\setlength{\skip\footins}{0.8cm}
\setlength{\footnotesep}{0.25cm}
\setlength{\jot}{10pt}
\titlespacing*{\section}{0pt}{4pt}{4pt}
\titlespacing*{\subsection}{0pt}{15pt}{1pt}

\fancyfoot{}
\fancyfoot[LO,RE]{\vspace{-7.1pt}\includegraphics[height=9pt]{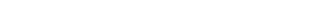}}
\fancyfoot[CO]{\vspace{-7.1pt}\hspace{13.2cm}\includegraphics{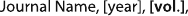}}
\fancyfoot[CE]{\vspace{-7.2pt}\hspace{-14.2cm}\includegraphics{head_foot/RF}}
\fancyfoot[RO]{\footnotesize{\sffamily{1--\pageref{LastPage} ~\textbar  \hspace{2pt}\thepage}}}
\fancyfoot[LE]{\footnotesize{\sffamily{\thepage~\textbar\hspace{3.45cm} 1--\pageref{LastPage}}}}
\fancyhead{}
\renewcommand{\headrulewidth}{0pt} 
\renewcommand{\footrulewidth}{0pt}
\setlength{\arrayrulewidth}{1pt}
\setlength{\columnsep}{6.5mm}
\setlength\bibsep{1pt}

\makeatletter 
\newlength{\figrulesep} 
\setlength{\figrulesep}{0.5\textfloatsep} 

\newcommand{\topfigrule}{\vspace*{-1pt}%
\noindent{\color{cream}\rule[-\figrulesep]{\columnwidth}{1.5pt}} }

\newcommand{\botfigrule}{\vspace*{-2pt}%
\noindent{\color{cream}\rule[\figrulesep]{\columnwidth}{1.5pt}} }

\newcommand{\dblfigrule}{\vspace*{-1pt}%
\noindent{\color{cream}\rule[-\figrulesep]{\textwidth}{1.5pt}} }

\makeatother

\twocolumn[
  \begin{@twocolumnfalse}
{\includegraphics[height=30pt]{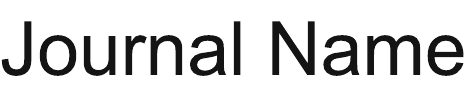}\hfill\raisebox{0pt}[0pt][0pt]{\includegraphics[height=55pt]{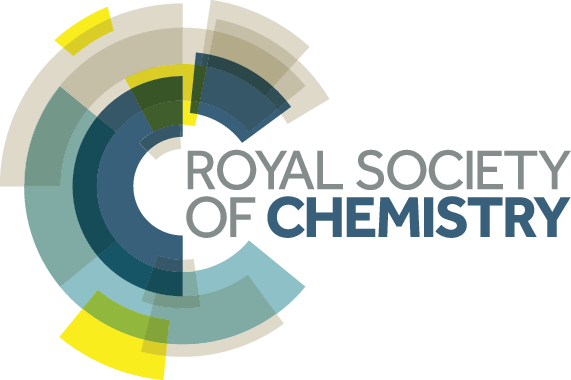}}\\[1ex]
\includegraphics[width=18.5cm]{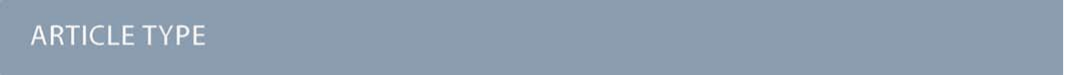}}\par
\vspace{1em}
\sffamily
\begin{tabular}{m{4.5cm} p{13.5cm} }

\includegraphics{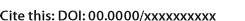} & \noindent\LARGE{\textbf{Quantifying Intuition: Bayesian Approach to Figures  of Merit in EXAFS Analysis of Magic Size Clusters}} \\
\vspace{0.3cm} & \vspace{0.3cm} \\

 & \noindent\large{Lucy Haddad,\textit{$^{a,b}$} Diego Gianolio,\textit{$^{b}$} David J. Dunstan, \textit{$^{a}$} and Andrei Sapelkin\textit{$^{a}$}} \\

\includegraphics{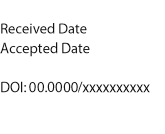} & \noindent\normalsize{Analysis of the extended X-ray absorption fine structure (EXAFS) can yield local structural information in magic size clusters even when other structural methods (such as X-ray diffraction) fail, but typically requires an initial guess -- an atomistic model. 
Model comparison is thus one of the most crucial steps in establishing atomic structure of nanoscale systems and relies critically on the corresponding figures of merit (delivered by the data analysis) to make a decision on the most suitable model of atomic arrangements. However, none of the currently used statistical figures of merit take into account such significant factor as parameter correlations. Here we show that ignoring such correlations may result in a selection of an incorrect structural model. We then report on a new metric based on Bayes theorem that addresses this problem. We show that our new metric is superior to the currently used in EXAFS analysis as it reliably yields correct structural models even in cases when other statistical criteria may fail. We then demonstrate the utility of the new figure of merit in comparison of structural models for CdS magic-size clusters using EXAFS data.  
} \\

\end{tabular}

 \end{@twocolumnfalse} \vspace{0.6cm}

  ]

\renewcommand*\rmdefault{bch}\normalfont\upshape
\rmfamily
\section*{}
\vspace{-1cm}


\footnotetext{\textit{$^{a}$~QMUL, Mile End Road, London E1 4NS UK; Tel: ; E-mail: apw813@qmul.ac.uk}}
\footnotetext{\textit{$^{b}$~Diamond Light Source, Diamond House Harwell Science \& Innovation Campus, Didcot OX11 0DE, UK. }}

\footnotetext{\dag~Electronic Supplementary Information (ESI) available: See DOI: 00.0000/00000000.}




\section{Introduction}
 
Establishing the atomic structures of  materials is a fundamental step in understanding their mechanical, electronic and optical properties and is essential for material applications. However,  recovering the atomic structures of nanomaterials is particularly challenging using standard structural analysis techniques (e.g. X-ray and electron diffraction, Raman scattering, etc.) due to loss of periodicity at atomic level and potential presence of novel metastable atomic arrangements. This is especially true of the recently discovered ultra-small truly mono-disperse nanoparticles --- magic-size clusters (MSCs) \cite{CdSe_MSC_Kui}, \cite{CdSe_pawthway_Kirkwood}, \cite{CdSe_Kudera}. As a consequence, several advanced structural methods such as X-ray absorption spectroscopy (XAS) and pair distribution function (PDF) analysis have been utilised to investigate atomic structure of MSCs \cite{MSC_Zhang},\cite{MSC_Lei_Tan},\cite{MSC_Ying}. XAS, in particular, has been shown to be sensitive to the atomic arrangements and structural changes in MSCs delivering information about sample stoichiometry and cluster symmetry discriminating between variety of structural models \cite{MSC_Ying}. 

\begin{figure}[h]
    \centering
    \includegraphics[width=0.4\textwidth]{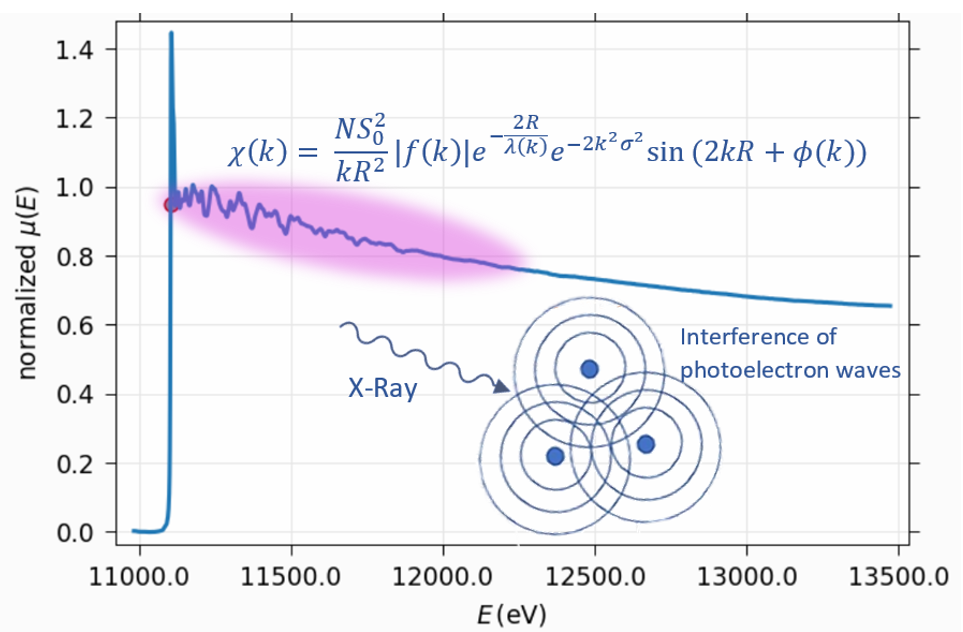}
    \caption{EXAFS analysis is the study and interpretation of the fluctuations in the post-edge X-ray absorption spectrum. The fluctuations in the signal (in the purple highlight) are the result of interference of the outgoing photoelectron wave with the portion scattered by the neighboring atoms. The EXAFS equation used for modelling these oscillations. $\chi(k)$ is related to the plotted absorption $\mu(E)$ by the transform: $ k=\sqrt{\frac{2m_e}{\hbar^2}(\hbar\omega-E_0)}$, $E_0$ being absorption edge energy.}
    \label{fig:firstexafs}
\end{figure}

This new class of nanoscale systems pushes XAS capabilities to the limit both in terms of the quality of the data required and of analysis methods for the two key parts of the X-ray absorption spectrum: X-ray absorption near edge structure (XANES) and extended X-ray absorption fine structure (EXAFS). The former is sensitive to the symmetry around the absorbing atom of interest (e.g. Cd in CdS MSCs \cite{MSC_Ying}) and its oxidation state, while the latter provides information about local coordination numbers, interatomic distances and local atomic dynamics (see Fig. \ref{fig:firstexafs}). 

Analysis of EXAFS data typically involves background subtraction and normalisation followed by theoretical EXAFS calculations for a selected structural model (or a selection of model structures), comparisons of the calculated spectrum with the data and parameter refinement to obtain the best fit and the corresponding structural information \cite{quantitativeRavel} \cite{crg2007introduction}. Theoretical calculations and subsequent refinement are some of the most crucial steps and require a suitable atomistic model, thus implying some prior knowledge of the atomic structure or having an informed guess (e.g. based on molecular dynamics, DFT calculations or similar material, etc.). Recovering atomic structure of MSCs puts particularly stringent demands on model comparison in EXAFS because local atomic arrangements can be quite similar \cite{airss}.
When theoretical and experimental spectra are compared, in most EXAFS analysis programs (such as Artemis \cite{Demeter} and Larch \cite{XRAYLARCH}) there are a number of figures of merit (FoMs) available to provide quantitative model evaluation to answer the question of whether the model is a suitable match for the experimental data. However, none of the commonly used FoMs take into account parameter correlation.
At the same time, it is well-known \cite{multicollinearity_LS},\cite{multicollinearity_1},\cite{multicollinearity_2} that correlation can have significant negative consequence on data refinement (i.e. larger errors) and, most importantly, on model verification and selection.

This shortcoming of the EXAFS FoMs has long been recognised as a problem and in the latest development of the Artemis (one of the most commonly used EAXFS analysis package) a heuristic "happiness parameter" is offered to provide in-code indication of the fit quality. This parameter is based on decades of EAXFS analysis experience and includes, with varying weighting, an R-factor (a numerical measure of how well the fit over-plots the data), penalties for parameter correlations, restraints, the number of independent parameters, etc. \cite{happiness}. While recognised as an important guide during data analysis, being a heuristic parameter, it has no firm basis in statistics and therefore cannot be quoted in publications. \\

In this article we introduce for the first time in EXAFS analysis an FoM that explicitly includes parameter correlations - the Bayes Factor Integral (BFI)\cite{BAYESDavid} 
We use EXAFS data for crystalline Ge at low temperature to demonstrate that the BFI is more sensitive than the typical FoMs used for EXAFS analysis to model choice.
We then demonstrate that the BFI consistently points towards the correct structure as preferred model.  We then use the BFI to compare a selection of models for a material with unknown structure: CdS Magic Sized Clusters (MSCs 311 and 322)  \cite{MSC_Ying} \cite{MSC_Lei_Tan} \cite{Dove_MSC}. With these examples, we introduce the BFI as a numerical metric for quantifying intuition in EXAFS model comparison. \\

\section{Methods}
\subsection{Figures of Merit in EXAFS analysis}

Least-squares fitting (LS, the minimisation of the sums of squares of residuals to optimise a model) is a commonly used method to fit data, to estimate parameters and to make decisions about model selection. In EXAFS LS fitting, reported FoMs in Larch and Artemis are $\chi^2, \chi_\nu^2$, R-factor, AIC and BIC.

The first is  the well-known statistical value characterising the residuals between the model and experimental data. It is a simple statistical measure of how small the fit residuals are: i.e. how closely does the model fit the data, However, it is has been well-established that the number of independent variables (fitting parameters) can significantly influence the value of $\chi^2$. The total number of parameters available in EXAFS analysis is limited by the sampling theorem of the Fourier analysis \cite{Nind} (this is also known as the Nyquist criterion/theorem in EXAFS community). Therefore, the most commonly reported fitting statistic in EXAFS is the so-called reduced chi-squared, $\chi_\nu^2$, based on $\chi^2$ but with a modification to include  normalisation of $\chi^2$ by degrees of freedom such that once the maximum number of free parameters allowed for the data ($N_{ind}$) is reached \cite{Nind}, it will become negative and provides a clear indication of over-fitting. The R-factor is another variation of the $\chi^2$ criterion with a different normalisation factor. \\

The AIC and BIC are not found in Artemis, but are used as FoMs in Larch \cite{XRAYLARCH} to aid model comparison: both are based on the Likelihood function (rather than $\chi^2$
) while also including a penalty term for adding parameters to the model (adding fitting parameters to a model --- physically meaningful or not --- normally increases the likelihood of the model while reducing the probability that the model is correct). \\

There are a number of problems with the figures of merit described above.  They treat all parameters alike, whether physically-meaningful or not.  Apart from the number of parameters approaching $N_{ind}$, there is not much help from these FoMs to tell whether one has a physically meaningful fit. Crucially, none of them include parameter correlations, while it is well documented \cite{multicollinearity_LS},\cite{multicollinearity_1},\cite{multicollinearity_2} that parameter correlations indicate over-fitting and have significant consequences on the refinement errors and model selection. For example, in EXAFS analysis it is well-established that there exist correlations between fitting parameters even when $N_{ind}$ \cite{Nind} is not exceeded \cite{E0paper},\cite{quantitativeRavel},\cite{XRAYLARCH} ,\cite{EXAFS_PD_Analysis},\cite{EXAFS_SIMS_STRUCTURE}. Both Artemis and Larch do provide functionality to calculate parameter correlations, but these are almost never used to assess the quality of the fit nor to aid model justification or selection. To compensate for that and to help guide users during the refinement in Artemis there is an inbuilt FoM that does include correlations: the Happiness parameter \cite{happiness}. However, it cannot be reported in publications since it has no mathematical basis: it is an  empirical FoM that can be adjusted between fits to accord with the user's preferences. Hence, there is a need for a FoM rooted in statistics that does include parameter correlations.\\

\subsection{Bayes Theorem, Bayes Factor and Bayes Factor Integral}

The goal of EXAFS analysis can be described as ``to find the best model parameters that fit the data'' or, more generally, ``to select the best model that fits the data''. This lends itself naturally to the Bayesian statistical analysis and the use of Bayes theorem:
\begin{equation}
\label{EqBayes1}
    P(M\vert D)=\frac{P(M)P(D\vert M)}{P(D)}
\end{equation}
where $P(M\vert D)$ represents the conditional probability of the model $M$, given the data  $D$. In the case of multi-parameter fitting (including EXAFS) this can be rewritten as (see, for example \cite{MacKay_occam}): 
\begin{equation}
\label{EqBayes2}
    P(\mathbf{w}\vert D, M)=\frac{P(D\vert \mathbf{w}, M)P(\mathbf{w}\vert M)}{P(D\vert M)}.
\end{equation}
 where $\mathbf{w}$ is the vector of parameter values. Models can then be compared by, for example, taking a ratio of their conditional probabilities $P(\mathbf{w}\vert D, M)$. This ratio of probabilities of two models (e.g. $i$ and $j$) represents the odds ratio in favour of one model over the other \cite{gregory_2005}:
 \begin{equation}
 \label{EqOdds}
     O_{ij}=\frac{P(M_i\vert D, I)}{P(M_j\vert D, I)}
 \end{equation}
 where $I$ is prior information we have about the models. Using Eq. \ref{EqBayes2} it is straightforward to show that:
 \begin{equation}
   O_{ij}=\frac{P(M_i\vert I)P(D\vert M_i, I)}{P(M_j\vert I)P(D\vert M_j, I)} =\frac{P(M_i\vert I)}{P(M_j\vert I)} BF_{ij}
 \end{equation}
 where $BF_{ij}$ is the \textit{Bayes factor}\cite{gregory_2005}. The ratio on the right hand side is the prior odds ratio of the two models and throughout this work we consider this to be unity (i.e. no preference of one model over another). Thus, to compare the models we need to compute $P(D\vert M_i, I)$ --- probability of the data given the model and prior information. However, expressions for $P(D\vert M_i, I)$ can be rather complicated and for analysis involving many (in general correlated) parameters they include the evaluation of a multi-dimensional integral over the parameter space (the MLI -- marginal likelihood integral). Assuming uniform top-hat priors and a Gaussian error distribution for independent identically distributed experimental data points gives (see for example \cite{gregory_2005}, p. 276):
 \begin{equation}\label{EqMatrix}
     P(D\vert M_i, I)=\frac{1}{\prod_{i=1}^{m}\Delta p_i}\frac{1}{\sigma^n(2\pi)^{n/2}}\int_{\Delta p}d^m \mathbf{p} e^{-\frac{1}{2}{\mathbf{p^T}\mathbf{Cov_p}^{-1}\mathbf{p}}}.
 \end{equation}
 where $\Delta p_i$ are prior parameter ranges, $n$ is the number of the data points, $m$ is the number of the model parameters $\mathbf p= [p_1 p_2 ... p_m]^{{\text T}}$ and $\mathbf{Cov_p}$ is the parameter covariance matrix.
Thus, although the Bayesian approach has already been demonstrated in application to EXAFS analysis \cite{viper}\cite{Krappe_Bayes}, it has not been used to any significant extent, as far as we can tell, on account of its complexity. Indeed, parameter correlation is almost always the case in EXAFS and would normally require evaluation of the multidimensional integral in Eq. \ref{EqMatrix}. That can be addressed by constructing an orthonormal set of the model basis functions \cite{gregory_2005} (model parameters) so that the new parameters will have no correlation and hence the multidimensional integral can be replaced by the product of multiple single integrals. However, this would require redefining the problem in terms of the new (orthogonal) parameter set, repeating the fit and then back-transforming the new parameters to recover the original ones. 
 
Here we propose a simple alternative FoM that requires only trivial modifications to the statistical procedures already existing in EXAFS analysis. We note that the multidimensional integral on the far right of the Eq. \ref{EqMatrix} constitutes the volume in the parameter distribution space. We also note that $\mathbf{Cov_p}$ is symmetric positive definite, hence its diagonalization involves basis rotation. However, the volume (and also $\text{det} (\bf{Cov_p})$) does not change under rotation of the parameter space required for the transformation to the orthonormal basis. Hence, the following expression normally corresponding to the orthonormal parameter set can be used for calculation of $P(D\vert M_i, I)$ --- \textit{Bayes Factor Integral (BFI)} --- for a model with parameter correlations present \cite{BAYESDavid}:
 \begin{equation}\label{EqBFI}
    BFI = (2\pi)^{m/2}L_{\text{max}}\frac{\sqrt{\text{det} (\bf{Cov_p})}}{\prod_{i=1}^{m}\Delta p_i}
\end{equation} \\
where $\Delta p_i$ are the initial parameter ranges and $L_{\text{max}}$ is the likelihood for the model. Thus defined $BFI$ can then be used for model comparison (giving preference for a model with the larger value of $BFI$) following EXAFS data fitting without the need to redefine the problem in the new orthonormal parameter set. We call this $BFI$ (rather than, for example, MLI)  to distinguish from a more common case when the orthogonal parameter set is used to obtain Eq. \ref{EqBFI} (and therefore the $\bf{Cov_p}$ is a diagonal matrix). Crucially, the FoM in Eq. \ref{EqBayes2} naturally incorporates the Occam factor:
\begin{equation}\label{EqOccam}
    \Omega = (2\pi)^{m/2}\frac{\sqrt{\text{det} (\bf{Cov_p})}}{\prod_{i=1}^{m}\Delta p_i}
\end{equation} \\
that accounts for parameter correlations as well as parameter ranges and provides a penalty for a model with significant parameter correlations and/or large initial parameter ranges (parameter uncertainty). However, since values of $BFI$ can vary drastically, a more convenient way of evaluation is through comparison of $\ln(BFI)$ of the corresponding models. In such a case model evaluation is reduced to calculation of the $\ln(BFI)$ ratios --- designated in this paper as $\ln(BF)$ ---  with the following scale\cite{BAYESDavid} \cite{Kass1995BayesFA} for $\ln(BF)$ values that differentiate between the models: 
\begin{itemize}
    \item $<1$ – barely worth considering,
    \item $1-2$ – substantial,
    \item $2-5$ – strong evidence,
    \item $>5$ – decisive.
\end{itemize}
We proceed below by testing this approach on a reference data set (crystalline Ge, c-Ge) before applying the procedure to the Magic Size clusters of CdS.

\section{Results and Discussion}

\subsection{The Case of Crystalline Germanium}

\begin{figure}
\centering
\begin{subfigure}{0.4\textwidth}
    \includegraphics[width=\textwidth]{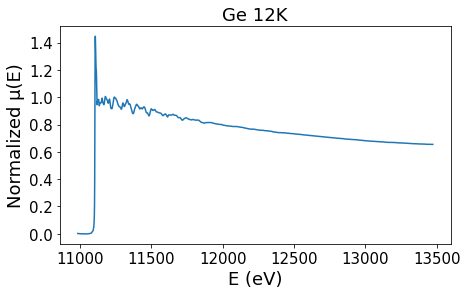}
    \caption{Ge spectrum in E-space.}
    \label{fig:ge_e_plot}
\end{subfigure}
\hfill
\begin{subfigure}{0.4\textwidth}
    \includegraphics[width=\textwidth]{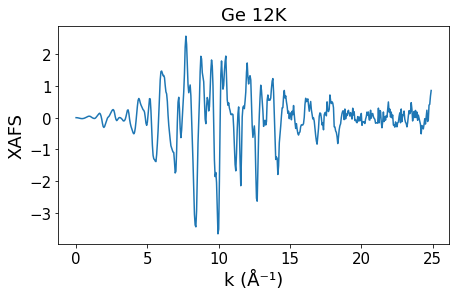}
    \caption{Ge spectrum in k-space.}
    \label{fig:ge_k_plot}
\end{subfigure}
\hfill
\begin{subfigure}{0.4\textwidth}
    \includegraphics[width=\textwidth]{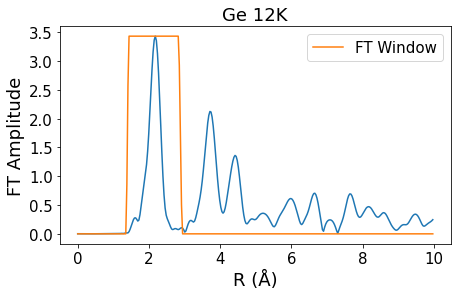}
    \caption{Ge spectrum in R-space.}
    \label{fig:ge_r_plot}
\end{subfigure}  
\caption{The Ge 12K XAS spectrum used for fits.}
\label{fig:E space spectrum}
\end{figure}

To verify the utility of BFI we first used the XAS data \cite{distancedwf} for Germanium collected at 12K (the x-ray absorption spectrum is shown in Figure \ref{fig:E space spectrum}). The data were selected on the account of their high quality.
The structure of Ge at this temperature is well-established and has been verified by previous publications \cite{Ge6fold}\cite{Gehighpressure}\cite{GeSapelkin}\cite{distancedwf}.The data analysis was performed in Larch \cite{XRAYLARCH} for background removal, normalisation and the actual fitting of the models to the experimental data since we found Larch to provide the most comprehensive fitting statistics. Figure \ref{fig:firstexafs} shows the EXAFS equation used to fit data where $N$ is number of nearest neighbours, $R$ is absorber-scatterer atom distance, $S_0^2$ is an amplitude reduction factor, $\sigma^2$ is Debye-Waller factor, $F(k)$ is photoelectron scattering amplitude, $\lambda(k)$ is photoelectron mean free path, and $\phi(k)$ is the phase shift. The latter three parameters ($F(k)$, $\lambda(k)$ and $\phi(k)$) are calculated using the FEFF 9 code \cite{xafs_theory_rehr} \cite{feffguide} and therefore are not refined.\\

Three structural models have been selected for comparison: (i) the actual structure of crystalline Ge at 12 K known to be a 4-coordinated face-centered diamond cubic type \cite{Gehighpressure} \cite{gehighpressure2} (Figure \ref{fig:n4cif}, $Fd3m$, Model 1); (ii) 6-coordinated high-pressure Ge phase VI structure ($Cmma$, Model 2); (iii) 6-coordinated high pressure $\beta$-Sn structure of Ge \cite{BetaSnGe} ($I4/amd$, Models 3a, 3b). For Model 3 two different refinements were carried out: one (Model 3a) was for a single shell of 6 nearest neighbours, while for the other (Model 3b) 3 shells of 2 atoms were used to reflect the actual nearest neighbour configuration in the $\beta$-Sn structure. This was also used to gauge the effect on the BFI of increasing the number of model parameters. 

To enable a fair model comparison, for each model we only looked at the first peak in the $R$-space (corresponding to the Ge-Ge bond length of 2.45 {\AA}; the atomic shell structure beyond the first shell is very different in the three selected models) and we used single-scattering paths only (see Figure \ref{fig:E space spectrum}). 
The data were fitted over the range of $2.00 < k < 22.93 \AA^{-1}$ in $k$-space (see Fig. \ref{fig:E space spectrum}). This ensured that only the first-shell EXAFS were fitted.   Parameter ranges are given in Table \ref{table1} and are defined as follows. The amplitude reduction factor $S_0^2$ corrects for inelastic effects in the absorbing atom \cite{bunker_2010}. This is empirically established to be in the range 0.8-1, and is well-covered by 0.5 range. The shift in the edge position $E_0$ accounts for errors in experimental calibration and for empirical convention in determination of the absorption edge position \cite{E0paper} \cite{quantitativeRavel} --- the range typically does not exceed 10 eV. Relative change in the nearest-neighbor interatomic distance $\Delta R$ is not expected to exceed 10\% as the interatomic distances are determined by the covalent radii of elements and the pressure-temperature conditions (as an example, 10\% bond length variation is well above that expected on melting or under pressures as high as 10 GPa in a typical semiconductor material such as Ge \cite{GeSapelkin}). Mean squared relative displacements of atoms due to atomic vibrations $\sigma^2$ (and static disorder, if any) accounts for damping effects on $\chi(k)$. The initial value can be calculated using e.g. correlated Debye or Einstein approximations \cite{Jeong2002LatticeDA}\cite{SevillanoDWF} \cite{dwf_ge_si} and for c-Ge at 12 K  this is around 0.003 \AA$^2$ \cite{distancedwf} hence the range of $\pm 0.003$ is selected to make it positive-definite. The number of nearest neighbours $N$ was set according to the structural models and was not refined.

\begin{table}[h!]
    \centering
    \begin{tabular}{c c c}
    \hline
    Parameter & Initial value & Range \\
    \hline 
    $S0^2$ & 0.9 &  0.5 \\
    E0 & 0 & 10 \\
    $\Delta R$ & 0 & 0.1 \\
    $\sigma^2$ & 0.003 & 0.006 \\
    \hline
\end{tabular}
\caption{Parameter ranges for all models.}
\label{table1}
\end{table}

For Model 1 (zinc-blende structure), one single-scattering single-shell path was used to fit the spectrum. For Model 2 (the high-pressure $Cmma$ structure), the spectrum was fit with 3 single-scattering single-shell paths between (in total) 6 atoms in the first shell to describe the signal. For ($\beta$-Sn) Model 3a, one single-scattering single-shell path was used at first, and then  3 single-scattering first shell paths were used (model 3b).\\

\begin{table}[h]
    \centering
    \begin{tabular}{cc}
    \hline
    Model & Ln(BFI) \\
    \hline
    1 & -6.65 \\
    2 & -10.07 \\
    3a & -7.32 \\
    3b & -10.22 \\
    \hline
    \end{tabular}
    \caption{Models and Ln(BF) values for the different Ge models.}
    \label{tab:Ge_BFI_table}
\end{table}

\begin{figure}
    \centering
    \includegraphics[width=0.4\textwidth]{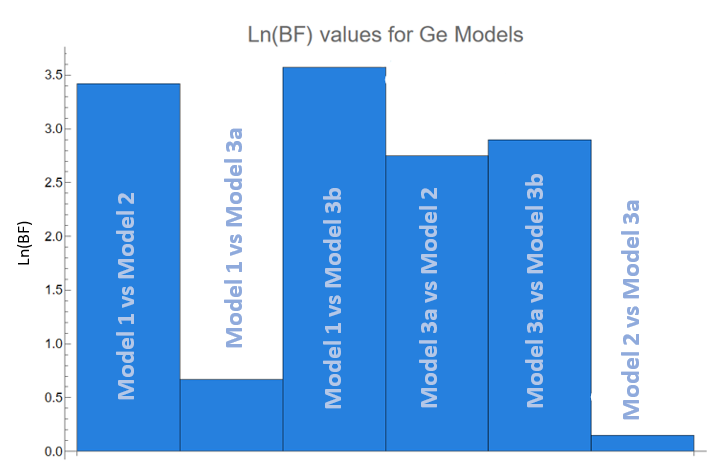}
    \caption{Ln(BF) values for the Ge models.}
    \label{fig:final_ge_compare}
\end{figure}

The summary of the results for the $\ln (BF)$ (the difference between $\ln (BFI)$ values) are shown in Fig. \ref{fig:final_ge_compare}.  One can see that Model 1 is favoured over all other models except for 3a (the single path $\beta$-Sn fit): the $\ln (BF)$ between Model 1 and all other models (except 3a) are found to be $>$ 3 providing strong evidence for Model 1 being the preferred structure. The $\ln BF$ value between Model 1 and 3a is 0.67 
is slightly in favour of Model 1 but not statistically significant according to the criteria outlined at the end of the previous section. However, the currently available fitting statistics FoMs found in the corresponding tables$^\dag$  favour other models: Model 2 has lowest $\chi^2$ and $\chi_{\nu}^2$, Model 3b has the lowest value of $R$-factor, while AIC and BIC favour Model 2 over Model 1. This shows that reliance only on the currently used FoMs in EXAFS analysis can lead to an incorrect atomic structure model as the best solution. At the same time, we see that the $BFI$ is able to deliver the correct result in this relatively complicated case -- after all we used a single peak only in the EXAFS FT magnitude in order to differentiate between the models. Having verified the utility of the proposed $BFI$-based FoM in case of the reference system,  in the next section we apply the Bayes approach to analysis of EXAFS of MSCs.


\subsection{Bulk CdS $k$-space Fitting}

Before proceeding on CdS MSCs we further tested the utility of the new FoM in $k$-space fitting of the bulk crystalline CdS. As reference data, bulk crystalline CdS EXAFS data at 90K at Cd K-edge were fit in Larch \cite{XRAYLARCH}. The first shell (Cd-S scattering paths) in k-space was fit using several different structures respectively: a zinc-blende, wurtzite \cite{WZ_CdS}, NaCl-like \cite{NaCl_CdS} and cmcm sturcutres \cite{CMCM_CdS}, latter two being high-pressure derived structures. \\

\begin{table}[h!]
\centering
\begin{tabular}{c c c}
    \hline
    Parameter & Initial value  & Range \\
    \hline
    $\Delta R$ & 0 & 0.1 \\
    E0 & 0 &10 \\
    $S0^2$ & 0.9 & 0.5 \\
    $\sigma^2$ & 0.006 &0.006 \\
    \hline
\end{tabular} \\
 \caption{Parameter values and ranges for all MSCs.}
  \label{table_MSC_params}
\end{table}

\begin{table}
\centering
\begin{tabular}{c c}
\hline
   Model  & Ln(BFI) \\
   \hline
    ZB & -0.738 \\
    WZ & -1.31 \\
    NaCl-Like & -2.04 \\
    cmcm & -2.31 \\
    \hline
\end{tabular}
\caption{Ln(BFI) values for each model in the bulk CdS EXAFS fitting (1st shell).}
\label{tab:CdS_Bulk_BFI}
\end{table}

K-space noise ($\epsilon_k$) was was evaluated from the signal between 6.50<k<18.30 {\AA} and the fit to the EXAFS data was carried out in the region 2.50<k<15.0 \AA, parameter ranges for the BFI calculations are shown in Table \ref{table_MSC_params}. The results of the fit are shown in the Table \ref{tab:CdS_Bulk_BFI}. The $BFI$-based FoMs support zinc-blende and wurtzite structures significantly over the NaCl-like and cmcm models. This is consistent with our previous results  where XPDF analysis of bulk CdS (and of regular CdS quantum dots) has shown CdS to be a mix of zinc-blende and wurtzite structures \cite{MSC_Lei_Tan}.

\subsection{Magic Sized Clusters}

Magic Sized Clusters (MSCs) are ultra-small ($<$3nm) colloidal semiconductor systems \cite{MSC_Zhang}. They are materials of interest due to their monodisperse nature \cite{CdSe_Kudera} that suggests one can deliver atomic-level control of system size using colloidal synthesis route. Their atomic structure is still under debate as are the methods of their structural verification. One of the key challenges for the latter is the possibility for stable and multiple meta-stable) atomic arrangements that are size- and temperature-dependent \cite{airss,MSC_Lei_Tan,MSC_Ying}.

The MSCs under investigation in this work are CdS. These MSCs exhibit a sharp UV-vis absorption peak at 311 nm (MSC 311) but when heated to ~$60^{\circ}$C \cite{MSC_Zhang} this peak shifts to 322 nm (MSC 322) \cite{MSC_Ying} and the shift is accompanied by atomic structure rearrangement as indicated by x-ray pair distribution function (xPDF) and XAS analysis \cite{MSC_Lei_Tan,MSC_Ying}. Due to their small size leading to the lack of long-range order, establishing the atomic-level structure of MSCs is challenging \cite{nanoproblems} and in this work we examine the sensitivity of our new EXAFS $BFI$-based FoM to the structural model selection. 
To this end we compared 4 models as candidates of possible structures of MSCs 311 and 322: i) Cd$_{40}$S$_{19}$ with zinc-blende structure, ii) Cd$_{40}$S$_{20}$ with wurtzite structure, iii) Cd$_{33}$S$_{32}$  $\beta$-Sn-like structure and ii) Cd$_{37}$S$_{20}$ with an  InP-like structure. The rationale for the model selection is as follows: 
\begin{enumerate}[i)]
\item
Bulk CdS can possess wurtzite or zinc-blende structure (while regular CdS quantum dots are known to exhibit both characters \cite{MSC_Lei_Tan}). Hence, when constructing a model for an unknown atomic structure of MSCs, the atomic arrangement found in the bulk can be a starting point if there is no other information (this is frequently the case).
\item 
It has been observed \cite{GeSapelkin} that average interatomic distances in small nanoparticles can be reduced compared to their bulk counterparts. This can be interpreted as an effective pressure on these systems. Such compression may results in distortion towards the  $\beta$-Sn structure \cite{whybetatin}, hence it is reasonable to use it as one of the structural models. 
\item 
It has recently been shown that an InP-like structure  \cite{InPsinglecrystal} 
provides the best fit to PDF \cite{MSC_Lei_Tan} 
data in CdS MSCs.
\end{enumerate}
All clusters have been cut as spherical regions of appropriate size from the corresponding bulk crystalline structure and were terminated with oxygen (except for InP-like cluster where the structure from our recent work \cite{MSC_Lei_Tan,MSC_Ying} was used). The bond length was set to that found in zinc-blende and wurtzite structure at ambient condition (2.55 {\AA}). After fitting with the initial structures they were all relaxed over 500 steps using Avogadro \cite{Avogadro} Universal Force Field \cite{UFF} and fitting process has been repeated. Data analysis has been carried out in Larch \cite{XRAYLARCH} and Mathematica 13.0 (for $BFI$ calculations). The initial parameter values and their ranges for all models are given in Table \ref{table_MSC_params} (if the standard errors returned by Larch exceed the half range set for $\Delta p_i$ terms, the range is corrected to 2 $\times$ standard error).

\begin{figure}
\centering
\begin{subfigure}{0.4\textwidth}
    \includegraphics[width=\textwidth]{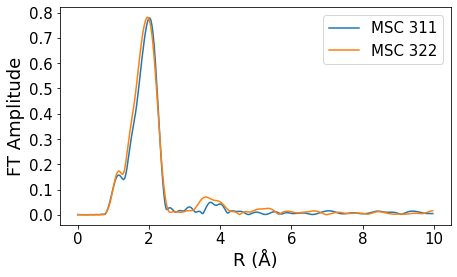}
    \caption{R-space EXAFS of the MSCs.}
\end{subfigure}
\hfill
\begin{subfigure}{0.4\textwidth}
    \includegraphics[width=\textwidth]{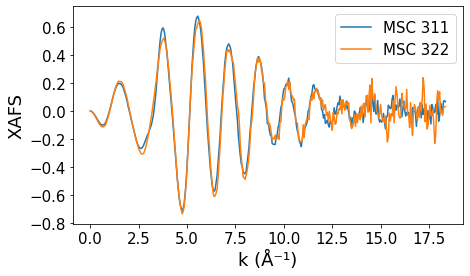}
    \caption{$k^2$-space EXAFS of the MSCs.}
\end{subfigure}
\caption{MSC 311 and 322 data used for fits.}
\label{fig:MSCdata}
\end{figure}

 
For each model the five highest-importance single-scattering paths were fitted up to the second cumulant ($\sigma^2$), we then fitted the third cumulant for the highest-importance Cd-O single scattering path of those included in the fit already.
While at low temperatures (90 K in our case) $3^{rd}$ cumulants for the atoms in the bulk of a nanoparticle are expected not to be significant, this may not be the case for the surface Cd-O coordination shell.  The $\ln (BFI)$s for and $\ln (BF)$s between the models were calculated using the same method as in the Ge model comparisons with the fits carried out in the range of $2-17$ \AA$^{-1}$ for the the coordination shell $1-3$ {\AA} (see Fig. \ref{fig:MSCdata}). For the noise value used in EXAFS chi-squared calculations we used the standard deviation of the $k$-space spectra for the MSC 311 and 322 data at $14.50-15$ \AA$^{-1}$ and $13-14$ \AA$^{-1}$ respectively.

\begin{figure}
\centering
\begin{subfigure}{0.4\textwidth}
    \includegraphics[width=\textwidth]{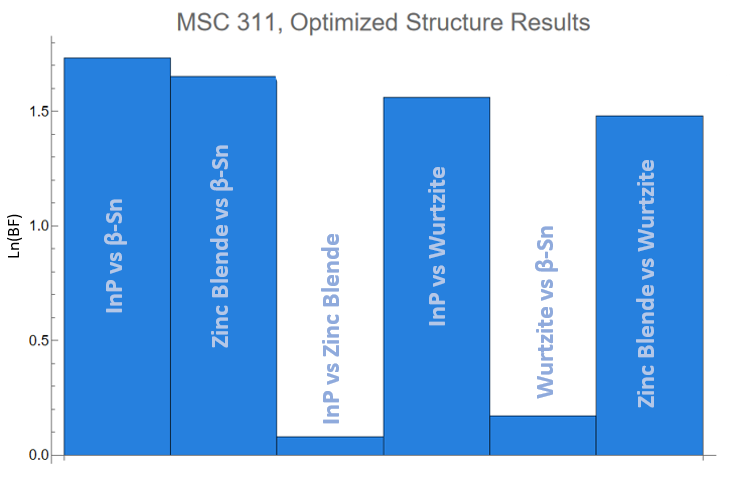}
    \caption{Ln(BF) values for MSC 311.}
    \label{fig:311comp}
\end{subfigure}
\hfill
\begin{subfigure}{0.4\textwidth}
    \includegraphics[width=\textwidth]{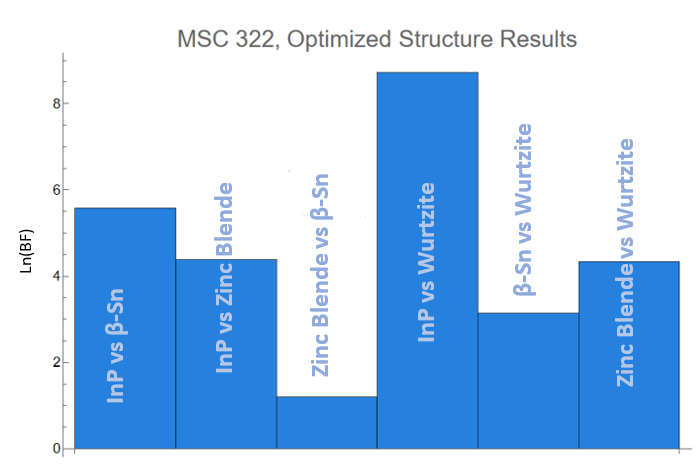}
    \caption{Ln(BF) values for MSC 322.}
    \label{fig:322comp}
\end{subfigure}
\caption{Charts of the results for each MSC.}
\label{fig:MSCBAR}
\end{figure}

The results for $\ln(BFI)$ for as-prepared and relaxed model clusters for 311 and 322 MSCs are given in the Tables \ref{tab:MSC311optimized} and \ref{tab:MSC322optimized} below and summarised in Figs. \ref{fig:311comp} and \ref{fig:322comp}.
\begin{table}[h!]
        \centering
    \begin{tabular}{c c}
       \hline
       Model & Ln(BFI) \\
       \hline
        zinc-blende & -5.71  \\
        $\beta$-Sn & -7.36 \\
        InP & -5.63 \\
        Wurtzite & -7.19 \\
        \hline
    \end{tabular}
    \caption{$\ln(BFI)$ values for optimized models, MSC 311.}
    \label{tab:MSC311optimized}
    \label{tab:msc_311}
    \end{table}
\begin{table}[h!]
        \centering
    \begin{tabular}{c c}
       \hline
       Model & Ln(BFI) \\
       \hline
        zinc-blende &  -10.78 \\
        $\beta$-Sn & -11.97 \\
        InP &  -6.39\\
        Wurtzite & -15.11 \\
        \hline
    \end{tabular}
    \caption{ln(BFI) values for optimized models, MSC 322.}
    \label{tab:MSC322optimized}
    \label{tab:msc_322}
    \end{table}
These results show that in the case of MSCs 311 InP and Zinc-Blende models are almost equally favoured over the $\beta$-Sn and Wurtzite structures. The results for $\ln(BFI)$ of 322 MSCs show that InP-like structure is significantly favored over all other models. Thus, our findings here indicate that the new FoM: i) is pointing to the InP-like model as the most probable structure; ii) is detecting the difference between the EXAFS data of the two MSCs (as reflected in changing values of FoM and model preferences). This is a very significant finding considering the differences between the EXAFS signals for MSC 311 and 322 are very small (see Figs. \ref{fig:MSCdata}) and the corresponding $\chi^2$-values are also quite close (see Supporting Information). The preference for InP-like structure is certainly in line with previous work \cite{MSC_Ying} \cite{MSC_Lei_Tan} where it was shown to provide the best model to fit xPDF, EXAFS and XANES data. All this provided very strong support to using the new Bayesian FoM we introduced in this work as a universal metric for model comparison and selection.


\section{Conclusions}

In this work we introduced the Bayesian-based statistical metric, the Bayes Factor Integral, for model comparison in EXAFS analysis. We showed for the first time that the new FoM provides a superior tool for model comparison in EXAFS by quantifying the intuitions about the parameter ranges and correlations through the Occam factor. We tested the new Bayesian FoM against reference EXAFS data for Ge and demonstrated that the it is superior to the FoMs typically used in EXAFS analysis and reliably predicts the correct structure. In the process we showed that ignoring model parameter correlations may result in a selection of an incorrect structural model. We then applied the new FoM for model comparison in analysis of MSCs where we demonstrated that it is sensitive to the differences between EXAFS signals of MSCs 311 and 322 and can point to the most probable structural models for these systems. 

So far we utilised identical parameter ranges for all models of interest in all our tests (except for the range correction when the standard error $\geq 1/2 \Delta p_i$). Imposing stricter model-specific parameter ranges in the calculation of model $BFI$ values should provide better results in model selection. Such model-specific parameter ranges can be obtained, for example, from molecular-dynamics or ab-initio simulations. 

We note that the current approach is so far limited by the requirement of providing initial guess structures, while the ultimate goal of structural analysis of MSCs (and of non-periodic systems in general) is producing structural models for materials with unknown structures. We believe that newly developed FoM can be the crucial aid on that pathway, for example by combining analysis described in this paper with machine learning/AI methods. \\

\section*{Author Contributions}

\section*{Conflicts of interest}
There are no conflicts to declare.

\section*{Acknowledgements}
LH is grateful to Diamond Light Source and Queen Mary University of London for the joint studentship and funding to support this work.



\balance


\bibliography{rsc} 
\bibliographystyle{rsc}

\appendix

\end{document}